# Buying time in software development: how estimates become commitments?

Patrícia G. F. Matsubara
Institute of Computing (UFAM) &
Faculty of Computing (UFMS)
patriciagfm@icomp.ufam.edu.br

Igor Steinmacher
Universidade Tecnológica Federal do
Paraná - Campus - Campo Mourão
igorfs@utfpr.edu.br

Bruno Gadelha
Institute of Computing (UFAM)
bruno@icomp.ufam.edu.br

Tayana Uchôa Conte
Institute of Computing (UFAM)
tayana@icomp.ufam.edu.br

*Abstract*— Despite years of research for improving accuracy, software practitioners still face software estimation difficulties. Expert judgment has been the prevalent method used in industry, and researchers' focus on raising realism in estimates when using it seems not to be enough for the much-expected improvements. Instead of focusing on the estimation process's technicalities, we investigated the interaction of the establishment of commitments with customers and software estimation. By observing estimation sessions and interviewing software professionals from companies in varying contexts, we found that defensible estimates and padding of software estimates are crucial in converting estimates into commitments. Our findings show that software professionals use padding for three different reasons: contingency buffer, completing other tasks, or improving the overall quality of the product. The reasons to pad have a common theme: *buying time to balance short- and long-term software development commitments*, including the repayment of technical debt. Such a theme emerged from the human aspects of the interaction of estimation and the establishment of commitments: pressures and customers' conflicting short and long-term needs play silent and unrevealed roles in-between the technical activities. Therefore, our study contributes to untangling the underlying phenomena, showing how the practices used by software practitioners help to deal with the human and social context in which estimation is embedded.

## I. Introduction

An estimate is a prediction about effort or costs [1]; a projection from the past to the future [2]. There is a degree of uncertainty involved. Nevertheless, in software projects, we use estimates for purposes that require precision to some extent. For instance, we use estimates in selecting or staffing projects, defining schedules, and generating quotes for customers [3]. Therefore, since different stakeholders interact to define, discuss, and agree on estimates, it is meaningful to understand how uncertain they are when communicating or receiving them. However, there is a large variance of what is meant by a software estimate, even inside the same company [4]. Estimators may provide an estimate they mean to be an ideal effort estimate to finish a task with almost no problems. Meanwhile, higher management may understand this same value as a risk-averse estimate. These distinct meanings are highly likely to go unnoticed by both parties [4]. In this scenario, a problem arises when establishing commitments regarding schedules or quotes for customers. The chances of attaining ideal effort estimates are much lower than risk-averse ones. This shows we cannot disregard human and social aspects regarding communication in estimation.

The problem, however, is not only a matter of communication and inconsistent terminology. The software engineering literature has warned us on the standard practice of assuming that our estimates and our goals are identical [5]. In other words, sometimes we mix our desirable business goals — or targets [1] — with our estimates. In some scenarios, this is imposed by external forces on the software teams, such as pressing real or virtual business needs or schedules' imposition. So, software professionals end up committing to a set of features associated with a schedule that their estimates do not support [6].

In other scenarios, mixing goals and estimates is unconscious and more subtle. For instance, in controlled experimental situations, where researchers simulated any perceived pressure associated with goals, research findings indicate that the knowledge of unrealistic client budgets led to lower estimates [7]. Therefore, in more realistic settings, with real customers involved and other pressing factors—like fierce competition, targets—are more likely to influence estimates, impacting their realism. Ultimately, the influence of desirable business outcomes on estimates leads to inaccuracies and errors. For instance, in bidding situations, there is a strong relationship between the focus on selecting developers with lower effort estimates and observed software project overruns [8]. Such overruns can significantly impact a company's profit levels.

To summarize, an estimate may indeed turn into a target or a commitment—a promise to deliver a feature at a certain level of quality by a specific deadline—but the opposite should not happen [1] for the sake of estimation accuracy and realism. Still, by having to establish commitments with management, customers (or customer representatives) could have a large impact on how software teams estimate their projects and tasks — and this goes beyond the estimation process's technicalities, entering the domain of the human and social context in which estimation is carried out. Thus, our research goal is to investigate the interaction between software estimation and the establishment of commitments in software companies.

We investigated this interaction in three sequential rounds of qualitative data collection and analysis in companies using expert judgment as their estimation method. We used the outputs of one round as input to the next one's design. In the first round of data collection, we spent 38 days—approximately 152 working hours—observing estimation sessions from two teams in one company. Starting with observations enabled us to identify the most critical issues about establishing commitments and estimations to explore them in-depth in the following rounds. Next, we interviewed team members from these two teams in addition to a third team in the same company in the second round of data collection and analysis. In the third round, we exercised our codes and categories in other contexts, interviewing software professionals from four other companies in different locations and business domains, furthering our understanding of the studied phenomenon.

Our results indicate that the need to establish commitments leads to the definition of defensible estimates during estimation sessions, in addition to getting consensus among estimators. There may be an emphasis on producing estimates that team leaders and managers can defend, leading estimators to adopt more optimistic estimates in some industry situations.

In other words, estimators may deliberately change their estimates because of the pressure. Another core finding of our study is the identification of the scenarios in which and reasons why software professionals use padding—an increase in estimates beyond the value the estimator honestly believes is needed to execute the software development task [9]. We found three reasons why software professionals pad estimates as part of establishing commitments with their customers. All these reasons share an underlying theme: buying time to balance different commitments. That is, software professionals use padding as a mechanism to manage situations in which short and long-term commitments conflict, helping professionals to ensure all of them are satisfied in the long run. Also, our results indicate the cases when padding is not possible due to short-term needs—like the customer has high expectations about one functionality, and it is strategic to deliver it quickly.

## II. Background

Even though past research on software estimation has focused on the creation and improvement of methods [10], software practitioners still face difficulties with inadequate software estimation in the industry [11], where the most prevalent and preferred method is expert judgment [12] [13]. The literature about expert judgment is usually concerned with accuracy, primarily attempting to increase estimates' realism. Researchers have investigated the impact of different factors over the estimates, including human aspects such as the estimators' (stakeholders responsible for the estimation) personality [14], role [15], and level of optimism [16]. Furthermore, research findings show that expert judgment estimates may be less realistic due to judgmental biases, like anchoring [17], framing effects [18], and sequence effects [19]. Despite all of this, factors affecting estimates still demand more attention from researchers [20].

Among the factors possibly affecting estimates, there is the issue of using estimates to establish commitments with management, customers, or customer representatives. External forces and pressing needs may lead software professionals to make commitments that their estimates do not support [6]. In this direction, Jørgensen [21] recommends distinguishing between the (i) PX effort estimate – that is, the estimator believes there is an X probability the value will not be exceeded; (ii) the planned effort, which is the effort used in the project plan; and (iii) and effort-to-win, which is the effort acceptable from the perspective of the market or the client. These terms help differentiate between estimation, planning, and bidding, which are processes for different purposes. Failing to make this differentiation hinders the realism of estimates. However, this distinction is not always clearly made in the industry, and when an executive asks for an estimate, what they may really want is a commitment tied to a desirable business outcome [1].

Regarding the role of customer expectations on estimates, Jørgensen and Sjøberg [22] found medium to large effect sizes in one experiment where researchers exposed participants to expected effort attributed to the projects' customers. Participants who received specifications with an exceptionally low number regarding expected effort had much lower estimates than participants who received high values. Jørgensen and Grimstad [7] also indicate that unrealistic client budgets' knowledge led to lower estimates.

The literature also reports on explicit pressures to change estimates due to customer or management expectations in industry. In the '90s, information system professionals reported that such pressure for increasing or decreasing estimates was associated with overruns [23]. Such results reverberate in the 2000s and 2010s, with reports about management and customer pressures leading to unrealistic estimates [24] and inaccuracies [25], intentional padding/shrinking [26], as well as estimates sometimes being a cave-in to people with more power [27].

The customer's selection strategy also impacts the estimates, and there is a strong relationship between the focus on selecting developers with lower effort estimates and observed overruns [8]. In an anecdote of a real-life situation, Halkjelsvik and Jørgensen [28] comment on a software company manager who reports that their company only wins contracts when they are overoptimistic about the time to complete the work. Nevertheless, they usually find out later that the estimate used as the basis for the price offer – their commitment – was too low. When applied to bidding scenarios, we call this the "winner's curse", which in high uncertainty situations and a high number of bidders may lead companies to have lower profit levels [29]. The winner's curse is a byproduct of the selection bias, which occurs when the client's selection process for providers leads to an over-representation of proposals based on overly optimistic estimates [8].

In our study, we further investigated the issues related to the interaction of the estimation process and the establishment of commitments in software companies, going beyond bidding situations. We explain how we executed this study to accomplish our research goal in the next sections.

## III. Research Methodology

To investigate the interaction between the software estimation and the establishment of commitments in software companies, we performed a qualitative study, which we detail in this section.

### A. Research Question

We defined the following research question for this study:

**RQ - How are software estimates used to establish software development commitments?** The focus is on identifying the interactions among stakeholders to define final estimation values, including the strategies for reaching an agreement on estimates. We also look for any strategies that software professionals use during the conversion of estimates into commitments.

### B. Studied organizations and participants

This research study had two phases. In the first phase, two companies agreed to participate. We selected one of them, referred to as Company A. The main criterion for selecting the company for this phase was: they must have at least one software team estimating their tasks regularly. We did not include the other company because their teams did not perform explicit estimation activities as part of their software process.

Company A is medium-sized, with over 100 employees, located in Campo Grande, MS – Brazil. They develop and maintain software for large telecommunication companies, working in close collaboration with their customers. They have four software development/maintenance teams, three of which participated in the study. Although Team A1 and Team A2 share the team manager and one senior developer, they have one dedicated team leader each. The team leaders also act as product owner and software analyst for their teams. Team A3 has a different organizational structure, with one team manager who also plays the team leader role. In A3, the software analyst/product owner is not the team leader. Additionally, all teams are composed of software developers and software testers.

The company adopted hybrid development methodologies, which is the norm in the software industry [30]. Several of their practices come from Scrum, like sprint planning, daily stand-up meetings, and sprint reviews. Nevertheless, they did not apply Scrum by the book entirely. For instance, regarding roles, the teams had specialists and were not cross-functional. Regarding practices, tasks were mostly self-assigned, but we also observed the team leader assigning tasks to specific team members. Masood, Hoda, and Blincoe [31] have reported these and other variations as part of Scrum in practice, emphasizing that some are not necessarily a misuse or abuse of the method.

Although Planning Poker is the most used estimation technique in agile software development [32], Teams A1 and A2 abandoned it after trying for a while. Their software teams were inexperienced and young, leading to slow justification rounds and long estimation sessions when using Planning Poker. They switched their technique. Their estimation sessions have the team leader's presence and at least one estimator representing team members for each of their primary software development activities: backend development, frontend development, and software testing. The estimators are not necessarily the team members allocated for the estimated tasks, although there is a high probability that they are. As we illustrate in Fig. 1, the estimation sessions encompass two main steps: gathering **individual estimates** from the estimators responsible for each activity in software development (Step 1) and deciding the team's **internal estimate** (Step 2), which is the value that the team commits to with the team leader. After the estimation session, the team leader converts the internal estimate to the **final estimate**, defining the team's **commitments** with the customer (Step 3 in Fig. 1).

Team A3 still uses Planning Poker for estimating. Interestingly, they report they also want to change their method for the same reasons Teams A1 and A2 did. All teams describe their user stories and tasks in cards and make them available to the whole team on Jira. Teams A1 and A2 also maintain a physical board for their tasks, updated daily. Additionally, all teams carry out the estimation of items and tasks immediately before the beginning of their sprints.

In the second phase of the study, we invited software professionals from other companies to understand our results in different contexts. We interviewed four practitioners from four other companies located in other cities. We selected such participants because their teams conduct software estimation activities regularly. All the participants reported that their companies use agile practices. All of them rely on expert-judgment for estimation, while none adopt Planning Poker.

These participants are from two companies, B and C, which are from Manaus, AM – Brazil. Company B works developing software solutions for a large multinational electronics company and employs over 1,000 people. Company C is a medium-sized company, with around 100 employees in total. They develop a wide variety of software solutions for companies in the electronics business, telecommunications, car dealerships, and others. Companies A, B, and C have in common the fact that they develop customized software solutions for specific companies.

Companies D and E develop subscription-based software for specific business areas. Their estimation process revolves around the launching of new functionality or new products. Company D is a large company in North Vancouver, BC – Canada, with 1,200 employees, developing software for the real estate business. Company E is in Curitiba, PR – Brazil, providing solutions for the financial area. It is a large-sized company, with around 700 employees.

### C. Data collection and analysis

Our study was conducted in three sequential rounds of data collection intertwined with data analysis: two rounds in Phase 1, focused on one company, and one round in Phase 2, expanding the research to other companies. In each round, we collected and analyzed data, using the output from one round to inform the next one's design, as Fig. 2 shows.

In the first round of data collection and analysis, we observed software estimation sessions and daily stand-up meetings [1] in Company A, as Table I shows. One of the researchers spent half-days in the company for 38 days from January to March of 2020. In total, this represents approximately 152 hours. Thus, the researcher was available to participate in their activities whenever there was the opportunity to do so. The researcher participated in all the estimation sessions during these days, taking notes of occurrences related to the research questions.

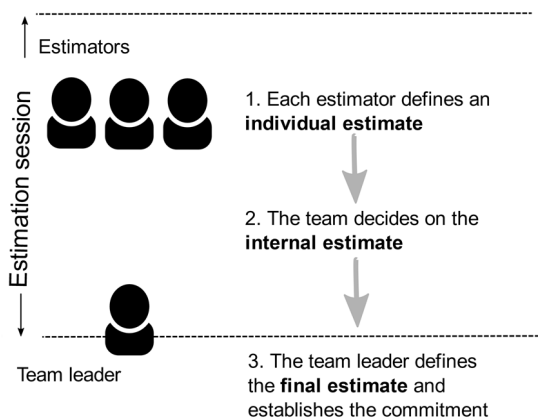

Fig. 1 – From estimates to commitments.

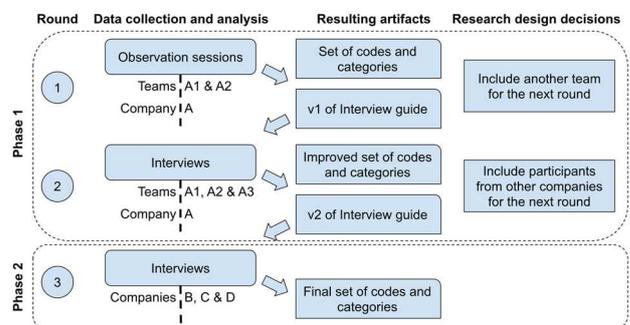

Fig. 2 – Phases and rounds of qualitative data collection and analysis.

TABLE I. OBSERVATION SESSIONS

| Team | Observation sessions | Participants |
|---|---|---|
| A1 | Three estimation sessions | Team leader, software developer, and software tester |
| A1 | 20 daily stand-up meetings | All A1 team |
| A2 | One estimation session | Team leader, software developer, and software tester |
| A2 | 20 daily stand-up meetings | All A2 team |

After observing the first estimation session, we analyzed the collected data, focusing on open coding [33]. We then proceeded with the other observation sessions. Again, we conducted open coding procedure to analyze the data. This round also resulted in an interview guide based on the main results from the analysis of the data from the observation sessions. For instance, we noticed that padding was a recurring theme in estimation sessions. For instance, we included the following question in our interview guide: [2] "In which situations do you add padding to software estimates?"

In the second round, we interviewed Team A1, A2, and A3 members, as shown in Table II, resulting in improvements to our set of codes and categories. We also changed our interview questions to focus more on disagreement resolution issues during estimating and on padding due to the analysis we had up to that moment. After the second round of data collection and analysis, we decided to investigate other contexts, moving to Phase 2 of the study. We expected that this would either confirm the results we found so far or lead us to discover more aspects about them. We proceeded to a third-round collecting data, in which we interviewed software professionals from the other four companies – represented in Table II as Companies B, C, D, and E.

TABLE II. INTERVIEW PARTICIPANTS[3]

| Interviewees | Roles | Company |
|---|---|---|
| P1, P2 | Team Manager | A |
| P3, P4, P5 | Team Leader, Product Owner, Software Analyst | A |
| P6, P7, P8 | Software Developer | A |
| P9 | Team Leader | B |
| P10 | Business Analyst, Software Analyst | C |
| P11 | Software Developer | D |
| P12 | Software Developer | E |

One of the researchers interviewed the participants, taking notes of their answers. At this point in the research, we intertwined data collection and analysis even more by coding each interview before proceeding to the next one.

During the data analysis, we created codes associated with the relevant parts of the annotations. The researchers held meetings to reach a consensus about the codes and ensuring they were grounded on data. We applied constant comparison throughout the analysis leading to the continuing evolution of the set of codes. We also discussed the relationships between the codes during the meetings as part of axial coding [33].

Finally, we presented our research results to participants of Teams A1, A2, and A3. They considered that the results were correct and reflected their current practice.

IV. RESULTS

This section presents our research findings regarding ***RQ - How are software estimates used to establish software development commitments?*** We present the codes and categories [4] – these last ones in bold – starting with the phenomenon of **defensible estimates**. Next, we move our attention to the **padding phenomenon**, a central part of converting software estimates into commitments. We explore the **padding scenarios** and **the reasons to pad**.

*A. Defensible estimates*

Since the teams we observed decided to abandon Planning Poker, their estimation sessions start with each participant providing their individual estimates. Following, the team defines their internal estimate. If they all agree on the individual estimates, then the value is set. However, they may face **disagreements**, leading them to adopt **disagreement resolution strategies**. For instance, the estimator may justify the given individual estimate as one step towards disagreement resolution, even though it may not be enough. If everyone accepts the justification, the proposed individual value is accepted. However, if they reject the justification, another step is to do another estimation session later. We also observed that when a disagreement occurs, they might set the estimate as an average of the proposed individual values or as the highest individual estimate.

Finally, another occurrence we observed in the face of disagreements is that the estimator changes their individual estimate. These changes happen in two situations: (i) when the other estimators strongly disagree with an individual estimate or (ii) when the team leader expresses that the internal estimate value is not defensible, which we illustrate with the following excerpt of an estimation session. In any case, the estimator moves to a more optimistic estimate.

> OP16[3]: "*I believe it takes three to four days to full development because for each period of the day developing for the web, I take two periods developing for the mobile platform.* (...)
> OP17, regarding software testing: "*It takes four days in total. Two days for local testing and two days for beta testing, because we have to evaluate the impact on System Y[5]*".

---

[1] We included the guiding questions for observation here: https://www.doi.org/10.6084/m9.figshare.13105319
[2] We included the interview scripts here: https://www.doi.org/10.6084/m9.figshare.13105319
[3] Participants we did not interview but who were involved in observation sessions were labeled as Observation Participants (OP): OP13-OP20.
[4] See more of our categories, codes and supporting quotes https://www.doi.org/10.6084/m9.figshare.13105319
[5] Names are omitted due to confidentiality issues.

> P4 made a totalization, registering it would take five days for backend development, plus five days for frontend development (in parallel with the backend), plus four days for testing - therefore, nine days in total;
> P4: "*I'll wait for the confirmation of the frontend development estimate. But you have to give it to me today.*" Next, thinking aloud: "*But I don't know whether I can defend nine days...*"
> So, OP17 answered that it could be one day and a half for each test type.
> Then P4 said: "*well... I can defend for eight days!*" - Estimation session from Team A1

The team leader considered the internal estimate not defensible at first. Then, one of the estimators changed his position – his individual estimate – to a more optimistic one to help the team leader to get to a defensible estimate. This occurrence led us to ask the team leader what makes an estimate defensible: "*Some estimates that software developers give me have too much padding, then I don't buy it. So, if they don't convince me, I won't be able to defend it. If they explain it to me during the estimation session and it makes sense, I accept it.*" (P4, Team Leader). Therefore, when the individual estimate is too high, it is not defensible. The team leader has some notion about the task complexity because the senior developer and the team manager give a baseline estimate for the task before the estimation session.

Nevertheless, individual estimates with explainable padding are acceptable, and discussing the solution options also contributes to the defense. The degree of novelty and the complexity are the task characteristics that explain an estimate, making it defensible, as the team leader continued to explain: "*So, a defensible estimate to me has a lot to do with the task complexity. I ask myself: are there many business rules involved in this task? Is there anything like this we have done before? If it is too novel or difficult, we must understand it and build the logic behind it with the team to inform the scope description.*" (P4, Team Leader).

To lower the pressure while maintaining a good relationship with the customer, the team leader devised a strategy to stand for the final estimates: detailing the items that make up the task and the estimate. P4 (team leader) talks about this: "*My customer is highly resistant to the deadlines I give. He tries to shorten all of them. The way I found to deal with this is to detail all the items of the estimate. This strategy is becoming our standard one, especially when the estimate is a little high because then the customer has no arguments. (...) A few days ago, I had registered a task on our tool, and the customer called me to talk about the deadline. When I informed him, he was like: "What?!" In these situations, I have to explain to him the estimate, showing item by item as I have registered in the tool, confronting them with the scope description (...) And I inform the deadline for each item. For more complicated tasks, I refine even further. So, the customer is accepting the deadlines I tell him. And this pressure is highly contingent on the customer*".

> **Takeaway message 1:** Apart from getting consensus from the team and making estimators committed to their estimates, estimation sessions also focus on building a defensible estimate.

*B. Padding scenarios*

Our results suggest that the padding phenomenon is essential in establishment of commitments with the customer. We identified three scenarios for the interaction between estimate and padding: (i) the estimator embeds padding in the individual estimate; (ii) the team leader adds padding to the internal estimate; and (iii) no padding at all is added to the estimate.

The first scenario is when **the estimator embeds padding in the individual estimate**. In this case, the estimator pads as part of giving their individual estimate. About this, P11 (software developer) states that: "*Every developer has their own estimation method. I believe we all add padding internally, but no one talks about it. As I am an optimistic fellow, I always pad, but I don't talk about it. If I think it takes one day, I will say it takes three. Some people may do it for slacking, but I do it because of my optimism since I have already had trouble giving lower estimates. Especially at the beginning of projects.*" P4 (team leader) also talked about it in one interview: "*The software developer wants to work without pressure. I receive lots of estimates with padding from them. It is rare to get an estimate of something to be done in one hour.*"

The second scenario is when **the team leader adds padding to the internal estimate**—i.e., to the estimate the team collectively agreed on—before committing with the customer. The following quote from P3 (team leader) shows it: "*So I take the team's estimates, and I add some padding - one or two days if the task is small and up to five days if the task is large - because the team is too inexperienced. (...) I always consider whether the person giving me the estimate is more optimistic or pessimistic. (...) In my team, we have a super optimistic fellow. So, we need to add more padding before giving the customer the estimate.*"

Also, the interviews with team leaders revealed that the presentation of estimates during the establishment of commitments requires care. Uncertain estimates are interpreted as single-point estimates. Approximate values are interpreted as padded estimates — and the customer tends to reject them. P4 explains it: "*Also, if I tell my customer the estimate is around fifteen days, he will assume it is exactly 15 days. P1 told me that in functionality Z he informed an estimate in the "around of" format, and it became a commitment.*" P3 also explains more about it: "*In this process, we realized that if we inform the customer deadlines with round numbers – like 10, 20 or 30 days – he always complains the value is high because he suspects we rounded the number. So, if the software developer said it takes 15 days, we inform the client it will take 17 days, for instance.*"

Another important finding of our study regarding such a scenario is **padding awareness**: the team leader sometimes conceals from the estimator the padding in the final estimate. P4 (team leader) discusses it: "*If the team tells me they are spending six days on it, I say we will spend more - within acceptable limits. And I do not tell the developers of the padding I added*". One software developer (P8) also revealed more about how he suspects the team leader adds padding, but software developers are not aware of it: "*I believe P3 [the team leader] adds padding later, but it is not of our concern. (...) During the meetings, they told us that the padding is to raise the confidence of the team leader with the customer.*"

Therefore, when the team leaders add padding to the internal estimates, they may be trying to raise the confidence that the team will keep commitments with the customer. However, they conceal the padding they added from the estimators, making estimators accountable for their individual estimate if the task is given to them. P8 (software developer) reinforces this: "*They [team leaders] don't convert the padding to the team - at least it is what they said in the meeting. If the developer estimates five days during the estimation session, he has five days to finish the task.*" Other reasons for concealing the padding from estimators are that high estimates may give the impression there is plenty of time to execute the task and lead to lower productivity.

However, there are situations when the estimators are aware of the padding the team leader adds. The transcription below shows an observation session where the team leader first says there is no room for padding for software development activities. Nevertheless, later during the meeting, he adds padding to the software testing activity.

> OP13 listed all the classes he remembered and concluded that the task would take at least one ideal day of work, depending on the person who will execute the task; OP14 agreed with him;
> OP15 said that in his opinion, it would take two days; OP13 reaffirmed that his estimate was contingent on the person executing the task;
> OP14 commented that they always pad a little. However, P3 said that this week there is no room for padding.
> OP20 estimated one day for testing. P3 said he would count one day and a half to test because of other stuff, which is also necessary to verify.
> In the end, the estimate for development only was one day and a half. - Estimation session from Team B

Overall, estimators are aware of padding when a similar task is complex, the task involves problematic parts of the system, or there is a need for more robust testing. In the specific case of the abovementioned estimation session, the team leader needs the padding to ensure that they will have enough tests – and the team leader makes it straightforward for the team that this is how they are going to use the padding. In doing so, the team leader is limiting the use of padding.

The third scenario is when **no padding at all is added to the estimate**. Different reasons cause this scenario during the establishment of the commitments with the customer: when the task is urgent, the task is simple, the task is noticeably clear to managers, or when the task has a pre-defined deadline. Additionally, when the customer has a high expectation over the task, it is also impossible to pad. The same happens when the task seems simple, although it is not, and the customer has some technical background. In one of the companies, their context requires technical expertise, which may not be available in their team. In this case, no padding is possible when there are qualified personnel to execute the task.

P3 (team leader) mentioned several of these reasons when asked about when padding is impossible: "*In situations where an outsider would consider the card to be simple, but the implementation is like "may God help us." For instance, module W's code is quite tricky because there is an impact in many other parts of the system. However, from the perspective of the customer, it's simple. (…) In urgent situations, it is also impossible to pad. Also, in tasks in which the customer has high expectations, we cannot make late deliveries. Those are the cards that lead us to overtime work. Another case is when the task is a promise from our board of directors. We receive them closed, with a defined deadline. (…) We also consider who the customer is because sometimes he has a technical background and will not accept padding, depending on the task. If the task is about labeling a field, he won't accept padding at all.*" Therefore, as the different padding scenarios show, there are tasks for which padding is not viable.

> **Takeaway message 2:** The use of padding varies across three scenarios: (i) the estimator embeds padding in the individual estimate; (ii) the team leader adds padding to the internal estimate; and (iii) no padding at all is added to the estimate.

### C. The reasons to pad

Our data also revealed that there are mainly three different reasons to pad: (i) padding for contingency buffer, (ii) padding for completing other tasks, and (iii) padding for improving the overall quality. We give an example of each of these reasons in Fig. 3, which we explain in the next paragraphs. The example is a simplification of reality since a real task's padding may involve all three reasons to pad. Our illustration makes a didactic separation of each reason. It also includes the example of a task with no padding at all – Task A – to aid in the development of our argument for one of the reasons to pad.

First, team leaders and estimators may use **padding for contingency buffer** to deal with risks that may occur during software development and maintenance of a task and raise the chances to fulfill deadline commitments. We illustrate this reason to pad in the case of Task B in Fig. 3, where padding was added to Task B's estimates to keep a reserve to deal with risks associated with this task. P8 (software developer) discusses it when questioned about why software developers give higher estimate values: "*Usually, it is because we are afraid of the problems we will have to face. Like, in larger tasks or tasks that involve implementation in some specific parts of the system, which have higher chances to have a problem there." The estimator also adds padding to their individual estimate, fearing accountability due to delays. It may also be the case that the estimator considers himself optimistic, or even because the task depends on another one executed by a teammate known to make deliveries with errors. Padding individual estimates may also happen when estimators have too many doubts regarding the task features, leading senior developers to consider problems during development.

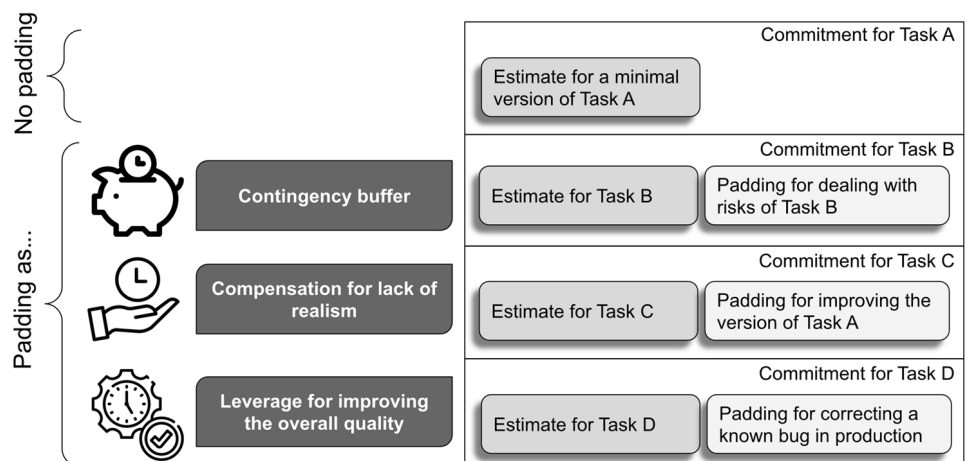

Fig. 3 – Example of reasons to pad.

Estimators also pad their individual estimates for more technical reasons, like the lack of familiarity with the company's code, if they have just begun a new job, as P11 (software developer) discusses: "*When I am at a new company, as I am not familiarized with their code, I add a high value of padding.*" Alternatively, they add padding when there are dependencies among tasks demanding lots of communication. Another more technical reason is when the task is related to problematic system modules.

Padding for contingency buffer is also useful when the team leader adds padding to the internal estimate when generating the final estimate. It may be used due to the tasks' characteristics, like when the task is large, critical, complex, or ill-defined. P9 (team leader) talks about this when asked in which situations he pads estimates: "*When we cannot define the feature very well. We need to carry our feasibility studies, but there is no time to do it because it is time to make a proposal*". It also happens when the implementation requires a learning curve or there is no time to investigate more about the task. Additionally, the team leader pads the internal estimate due to experience issues, like the team leader's past experiences or when the team is inexperienced. More technical issues may also play a role in padding for contingency buffer since it may happen because there is a need for integration with third-party software or the lack of technology specialists.

Specific issues related to the estimator's characteristics may also cause the team leader to add padding to the internal estimate, like when the estimator is inexperienced at the company. Another reason is a known higher level of optimism of the estimator, a known higher level of deliveries with errors from the estimator, or when the estimator is insecure. P3 (team leader) talks about it: "*Nowadays, I know when a task is going to return [with errors from the test] due to the experience I have with the person [assigned to the task] – then we add more padding. If the person is optimistic, we also pad.*" More generically, the team leader may pad for contingency buffer simply to deal with unforeseen problems or raise the confidence that the team will meet commitments. In any of these cases, dealing with risks seems to be essential.

The second reason for padding is **padding for completing other tasks**. It happens to gain time to implement a task that estimators or team leaders could not add padding for. We illustrate this case with Task A and Task C in Fig. 3. There was a need for padding to ensure the complete implementation of Task A, but the context did not allow for it. Therefore, the decision was to estimate Task A to deliver a minimally viable version of it and pad the Task C estimate to compensate. The padding of Task C is meant to be used to finish the full version of Task A instead of being used for Task C implementation. P3 (team leader) talked about this: "*We may use padding to gain time for a task that we could not add padding (...) We gave an estimate of 30 working days for functionality Y, but we are counting on the padding of other tasks to finish it.*" Notice this reason to pad connects with the scenario where there is no padding added to the estimate.

Padding for completing other tasks also happens when estimators or team leaders use padding from one task to implement tasks planned to, but not delivered in previous projects/iterations. P10 (business/software analyst) gave an illustrative example of this: "*Sometimes, we have a contract including functionalities A, B, C, and D, but we do not deliver D, for instance. So, we will implement D in another project, which includes other requirements, and we add padding for D*". In this case, the need for completing is also there, but the granularity is larger: the team needs to complete an entire project instead of a single module or functionality.

The third reason is **padding for improving the overall quality** of the product. It happens when team leaders use the value of padding when there is a need for more robust testing or to implement tests. It also happens when the team wishes to implement improvements in the system, or simply to develop carefully. Another motivation is to allow for the correction of bugs in production, as we illustrate in Task D of Fig. 3. Finally, padding for improving the overall quality may also happen to evolve well-accepted features. In other words, the estimator or the team leader can use the value added to the internal estimate to ensure the fulfillment of the established commitments with the customer, including overall quality commitments. P3 (team leader) talked about it in the interview: "*It also happens that there are errors we know, and we add one day in one task to correct it. For instance, we delivered functionality T, but we were not able to test it. After we delivered, the testers started to work, and they found lots of bugs. Now we are correcting these bugs.*"

On the one hand, it looks like team leaders use padding to meet short-term commitments, either by not padding tasks that, for instance, the client has high expectations or by including padding in estimates of tasks at hand to deal with risks. On the other hand, they also use padding to keep up longer-term commitments, like when they compensate for the lack of realism in some tasks through others' padding. The lack of time to dedicate to quality requirements in some tasks is also compensated through other tasks' padding, leveraging the product's overall quality. Therefore, they make sure to execute all tasks and satisfy all commitments in the long run.

> **Takeaway message 3:** We found three reasons for padding estimates: contingency buffer, completing other tasks, or improving the overall quality. These different uses of padding emerge from the estimation process to ensure both short and long-term commitments can be met in the long run, even when they are conflicting at a given moment of software development.

## V. Discussion

### A. Getting defensible estimates

One of our core findings is that estimation sessions serve to build defensible estimates, in addition to getting consensus among the team members and their commitment to estimates, as we presented in Section IV.A. This happens because team leaders may not be willing to accept estimates they are not convinced of. After all, they are the ones in contact with customers and who will negotiate with them to establish commitments based on these estimates. In response to this, estimators may explicitly change their initial estimates to more optimistic ones if their team leaders do not consider them defensible. However, changing estimates to more optimistic ones may lead to unrealistic estimates and errors.

The finding of the changes to estimates also aligns with the ones from an interview study with large and mature organizations, where Magazinius et al. report that estimators may decrease estimates due to management pressure or may change estimates to attain to organizational agenda, like due to the interests of customers [34]. These results show that, instead of standing up for their estimates and treating them as non-negotiable facts, technical staff still need to learn skills to convince their bosses of their estimates, as McConnell [35] said they were years ago. Thereby, it is not enough to provide estimators suitable methods for reaching consensus over an estimate or for generating a realistic estimate: it must be defensible.

> **Implications for practice:** To avoid pressure over their estimates, software professionals need to help their team leaders and managers to build up arguments for defending estimates during the establishment of commitments.
>
> **Implications for research:** Our results indicate that estimators change their estimates to more optimistic ones under pressure. Therefore, we need practices that empower such estimators to defend their estimates to keep them realistic.

*B. Padding estimates to buy time*

Another finding of our study concerns the padding phenomenon that we explored in Sections IV.B and IV.C, which involves adding a value to the original estimates before their communication when defining a commitment. We found industry scenarios in which padding is impossible, even if the team feels it is needed. When padding is viable, our findings indicate it is used to "buy" time for three reasons: for contingency buffer, completing other tasks, or improving the overall product quality.

Padding for contingency protects against risks in software development, buying time to deal with them. The use of contingency reserves for schedule and budget is already known as a recommendation for project management in the Project Management Body of Knowledge (PMBOK) [36], a good risk management practice to fight against fires that may impair a software project [5], and as a mechanism to compensate for the winner`s curse [28]. Also, the inclusion of a large buffer to deal with unexpected events or changes in specifications is a reason for accurate estimates [37]. Additionally, Magazinius et al. report that project stakeholders from industry sometimes intentionally increase their estimates to avoid overspending software development resources [34].

Therefore, the use of padding for contingency buffer is valid and vital for software tasks' execution – and it has been widely recognized in the software engineering literature. However, our results indicate that software professionals use padding for two additional reasons: completing other tasks and improving the overall quality. For completing other tasks, padding is added to one task to gain time to implement another one that they could not add padding for. The last task's commitment is not realistic, and the padding of other tasks counterbalances this fact. It also happens when a task was planned to be delivered in a given project/iteration but is not. Then, padding may be added to other projects/iterations to include these tasks. In any of these cases, padding serves for buying time for those other tasks.

Another reason to pad is for improving the overall product quality by implementing improvements in the system, amplifying tests, or allowing for the correction of bugs in production. It is like buying time to attain to quality requirements, satisfying long-run commitments. In alignment with our findings of padding for completing other tasks and improving the overall product quality, Magazinius et al. [34] report that the most common reason for intentional increases of estimates in their study was for hiding other activities in the estimated ones. They state this happens either to get more development time for one functionality or other testing or maintenance activities [34].

In such cases, padding is a managerial mechanism to allow for the repayment of technical debt in the software products. Lim, Taksande, and Seaman [38] report that management may not always recognize the importance of repaying technical debt unless they are rewarded or the customer is willing to pay for it. Additionally, customers may not be willing to give software teams the time to repay technical debt unless they get direct value from this [38]. Our findings indicate that in such a scenario, using padding to implement tasks not delivered in previous projects - padding for completing other tasks - is a way to repay requirements debt. Also, padding for implementing tests, implementing improvements in the system, or allowing for the correction of bugs in production - padding for improving the overall quality – is a mechanism to repay design, coding, or testing debts. Therefore, while researchers are focusing on more technical approaches for repaying technical debt, like refactoring, rewriting, automation, and others [39], industry professionals also have to find managerial paths to allow for such repayments, like padding their estimates.

Additionally, Becker, Walker, and McCord [40] mapped studies about intertemporal choices – a concept of psychology and behavioral economics referring to "decisions involving tradeoffs among costs and benefits occurring in different times" [41] – in software engineering. They found that no empirical work investigated trade-offs in time in depth. Our study contributes to filling in this gap, providing evidence about how practitioners use padding – or the lack of it – to balance short and long-term needs. Customers may have a strong focus on shorter time to delivery and lower costs, leading teams to sacrifice quality during software development. Such an attitude reflects on the estimation process, and the set of tasks at hand in a particular moment receives much attention. In this context, padding is a mechanism that team leaders and managers use for buying time to deal with risks in software development, to compensate for the lack of realism of previous tasks, or to improve the overall quality of the product in the long run.

Along with our results, the findings from these other studies indicate that padding is a relevant practice in the industry's estimating process, especially for protecting software projects from risks and providing managerial mechanisms for the repayment of technical debt. An interesting remark is that one of the team managers asked the researchers to present the research results regarding padding to a novice team leader for training purposes, which indicates their practical usefulness and relevance. It is time to recognize padding as another tool in the software engineers' toolbox to deal with estimation's social and human aspects.

> **Implications for practice:** Padding is a relevant practice in the software engineerings' toolbox and goes beyond providing a contingency buffer: it is also used to complete other tasks and improve the overall quality of a product. Practitioners can use our results to train novice team leaders on when and why to pad, given the reasons we found.
>
> Software teams can also classify their padding of tasks according to the reasons to pad. Too many tasks with padding for completing other tasks or improving overall quality suggest a need for improving – or perhaps defending – estimates.
>
> **Implications for research:** Padding hides the balancing of short and long-term commitments from customers. Sometimes a task is not padded to satisfy a short-term need – like delivering faster – but another one is padded to compensate for the resulting lower quality – like for correcting bugs left due to the absence of time for testing correctly. A better comprehension of padding in the software industry aids researchers in proposing alternative or supplementary practices to padding to make the balance of short and long-term commitments more transparent and controllable, instead of just yielding to the pressure of short-term needs.

## VI. LIMITATIONS

One of the limitations was that respondents might have understood interview questions differently from what we meant. To minimize this, we executed the observation sessions before the interviews to ensure we would use participants' terminology. By doing so, we also focused on specific behaviors closely related to our research questions.

Also, there was the risk some topics were too sensitive for participants to mention, as the changes of estimates and padding behaviors. For instance, in the research about distortions of software estimates, Magazinius et al. [34] comment on how some of their respondents asked them to stop audio recording in some parts of the interview to inform about sensitive issues. Likewise, we were running the risk of having our results biased by political reasons. We mitigated this risk by being in constant contact with the team and executing observation sessions for an extended period, making it unlikely that sensitive behaviors would be covered. We intended to promote an environment where participants could speak freely about any subject, including sensitive topics. Therefore, we did not audio record the observation sessions and interviews, raising the risk for misunderstandings. We showed a sample of our annotations to the participants of th observation sessions to validate them, in order to minimize this threat. After each interview session, we typed all the annotations and emailed them to the interviewee, asking him/her to read them and point inaccuracies. Additionally, we presented our results to some participants to assess their resonance — and we received positive feedback.

We first analyzed the data from one single company. Therefore, it was hard to say our results generalize to other contexts. We interviewed software professionals from other companies located in other cities and working in different business areas to address this. Despite the variation, all companies embrace agile or hybrid development to some extent. So, it may be the case that our results are especially relevant for this context. In any case, our main concern was with understanding a specific phenomenon over having generalizable results.

Concerning reliability, all the researchers held meetings for reaching a consensus during coding, ensuring that the codes were meaningful, representing the quotations and that the relationships between the codes were grounded on data.

## VII. CONCLUSIONS

This paper presented a study in five companies to investigate the interaction between software estimation and the establishment of commitments with customers. In this sense, our study contributes to untangling the underlying phenomena of defensible estimates and padding, showing how the practices software practitioners use in the field help them deal with the human and social context in which estimation is embedded. First, our results show that the interaction of estimation and the establishment of commitments lead to estimation sessions that focus on more than solely getting consensus among team members and making estimators committed to estimates. It also serves to build defensible estimates. Given this, estimators may change their estimates to more optimistic ones if there is a belief that they are not defensible.

Second, padding is a valid mechanism in the industry, and team leaders have different reasons to pad. They may use padding for contingency buffer, completing other tasks or improving the overall quality of the product. As a contingency buffer, padding serves as a reserve to deal with risks during software development and maintenance. For completing other tasks, the padding of one task embeds the estimates of other tasks for which padding was impossible. For improving quality, padding compensates for previous deliveries where time had higher priority over quality requirements. Interestingly, padding for completing other tasks and for improving quality are managerial paths that industry practitioners have found to repay technical debt.

Future work on defensible estimates involves developing mechanisms to provide software professionals the skills to defend their estimates, as McConnell [35] suggested years ago. Another possibility is proposing and investigating mechanisms that help software teams help team leaders build defensible estimates before the customer without sacrificing realism. Regarding padding, future work includes the proposing and investigating mechanisms to support appropriate padding, given each of the different reasons it is used for. Finally, software teams may benefit from alternatives to padding to help them to balance short- and long-term commitments in more transparent and controllable ways.

## ACKNOWLEDGMENT

This research, carried out within the scope of the Samsung-UFAM Project for Education and Research (SUPER), according to Article 48 of Decree no 6.008/2006(SUFRAMA), was funded by Samsung Electronics of Amazonia Ltda., under the terms of Federal Law no 8.387/1991, through agreement 001/2020, signed with Federal University of Amazonas and FAEPI, Brazil and through agreement no 003/2019 (PROPPGI), signed with ICOMP/UFAM. Also supported by CAPES - Financing Code 001, CNPq processes 314174/2020-6 and 313067/2020-1, and FAPEAM process 062.00150/2020. We also thank all the study participants.